# Phonon Spectrum Engineering in Rolled-up Nano- and Micro-Architectures

**Vladimir M. Fomin[1],* and Alexander A. Balandin[2]**

[1]Institute for Integrative Nanosciences (IIN), Leibniz Institute for Solid State and Materials Research (IFW) Dresden, Dresden, D-01069, Germany; E-Mail: v.fomin@ifw-dresden.de
[2]Phonon Optimized Engineered Materials (POEM) Center, Department of Electrical and Computer Engineering, University of California – Riverside, Riverside, California 92521 USA; E-Mail: balandin@ece.ucr.edu

*Author to whom correspondence should be addressed; E-Mail: v.fomin@ifw-dresden.de
Tel.: +49-351-4659-780; Fax: +49-351-4659-782.

**Abstract:** We report on a possibility of efficient engineering of the acoustic phonon energy spectrum in multishell tubular structures produced by a novel high-tech method of self-organization of nano- and micro-architectures. The strain-driven roll-up procedure paved the way for novel classes of metamaterials such as single semiconductor radial micro- and nano-crystals and multi-layer spiral micro- and nano-superlattices. The acoustic phonon dispersion is determined by solving the equations of elastodynamics for InAs and GaAs material systems. It is shown that the number of shells is an important control parameter of the phonon dispersion together with the structure dimensions and acoustic impedance mismatch between the superlattice layers. The obtained results suggest that rolled up nano-architectures have potential for thermoelectric applications owing to a possibility of significant reduction of the thermal conductivity without degradation of the electronic transport.

## 1. Introduction

Spatial confinement of acoustic and optical phonons in semiconductor thin films, superlattices and nanowires changes their properties in comparison with bulk materials [1-15]. Phonon confinement in nanostructures leads to emergence of the quantized energy subbands with corresponding modification of the phonon density of states [1-4, 10-15]. The changes in the phonon dispersion give rise to changes in the electron-phonon scattering rates [15-21], optical properties of the nanostructured materials [5, 22-28], and phonon scattering on defects, boundaries and other phonons [10, 12, 13, 29-31]. It was predicted that the electron mobility can be increased in the phonon engineered core-shell nanowires and planar heterostructures with the barrier shell materials, which are acoustically harder than the core materials [32, 33]. The experimentally measured reduction of the phonon thermal conductivity in thin films and nanowires usually results from the increased phonon – rough boundary scattering. Another mechanism of the thermal conductivity reduction is related to the phonon spectrum modification and decrease of the phonon group velocity in thin films [10, 29, 34] and nanowires [12-14, 30, 31, 34]. The predicted two-orders of magnitude decrease of the thermal conductivity in silicon nanowires at room temperature [31] has recently found experimental confirmation [35]. It has been determined that the thermal conductivity of the sub-20 nm diameter nanowires is suppressed by the phonon confinement effects beyond the diffusive boundary scattering limit [35]. The acoustic phonon confinement effects predicted theoretically for thin films and nanowires [6-34] have been directly observed experimentally using the Brillouin light scattering technique using suspended silicon thin films with the thickness H~7 nm [36] and gallium nitride nanowires with the diameters D~150 nm [37].

The initial work on the phonon confinement effects in thin films and nanowires was performed using the elastic continuum approximation [10, 11, 14, 29-31, 38]. The results obtained with the elastic continuum approach have been confirmed using other techniques such as lattice dynamics [34, 39] and molecular dynamics (MD) [40-42]. All computational approaches proved that the acoustic impedance $\eta = \rho \times v$ of the barrier shells of nanostructures presents an important tuning parameter for phonon transport, which can be used together with the lateral dimensions and shape for phonon engineering of material properties ($\rho$ is the mass density and $v$ is the sound velocity of the material). In the core-shell nanostructures with the acoustically mismatched barrier shells new types of phonon modes appear. Some of them are mainly concentrated in the nanostructure core while others are localized in the shell layers [11, 14, 34, 38, 39]. Controlling the acoustic impedance mismatch and thickness of the shell layers one can tune the electron – phonon interaction and the electron mobility [32, 33]. Similarly, one can engineer the phonon group velocity and thermal conductivity in such nanostructures [11, 14, 34, 38, 39].

Further progress in phonon engineering of material properties depends on availability of nanostructures with layered structures and substantial acoustic impedance mismatch. Multilayer tubes have attracted a special attention in newly developed acoustic metamaterials and phononic crystals at the micro- and nanoscale [43, 44]. In particular, a cylindrical structure from 40 alternating layers of 0.36 mm thick natural latex rubber film and 0.38 mm thick silicone elastomers containing boron nitride particles serves as a thermal shield [45]. In a two-layered tube with the weak interfaces between the layers, the dispersion characteristics of longitudinal guided acoustic wave provide a tool for detecting

4and exploring defects [46]. The microstructures are shown to play a decisive role in the dispersion of both flexural and longitudinal waves in single- and multiwall carbon nanotubes [47].

Acoustic metamaterials and phononic crystals are formed by periodic variation of the acoustic properties of the materials (elasticity and/or density), what leads to the occurrence of the phononic band gaps and provides powerful tools to control the phonon velocity spectra. The phonon crystals were proposed theoretically for elastic waves propagating in a composite material consisting of identical spheres [48] and infinite cylinders [49, 50] with parallel axes embedded in a periodic way within a host. First sonic materials with effective negative elastic constants were fabricated as lead coated spheres arranged in a simple cubic crystal [51]. They acted as total wave reflectors within certain adjustable acoustic frequency ranges. Split-ring resonator periodic arrays [52] and double negative (with the negative effective bulk modulus and the negative effective density) acoustic metamaterials [53] were suggested in a close analogy with electromagnetic metamaterials. For diverse experimental realizations of phononic crystals and their applications, see [44].

A novel method of self-assembly of micro- and nanoarchitectures was designed on the base of the strain-driven roll-up procedure [54, 55]. It paved the way for novel classes of metamaterials: single semiconductor micro- and nanotubes (or radial crystals) [56] and multilayer spiral micro- and nanotubes (or radial superlattices) [57]. A comprehensive structural study was provided for semiconductor/oxide, semiconductor/organic as well as semiconductor/metal hybrid radial superlattices [58]. A combined "roll-up press-back" technology has been recently presented to fabricate novel acoustic metamaterials – mechanically joined nanomembrane superlattices [59], which reveal a significant reduction of the measured cross-sectional phonon transport compared to a single nanomembrane layer.

The optical phonon spectra in multilayer cylindrical quantum wires manifest a geometric structural effect [4], which is of immanent importance for understanding of the pairing of charge carries in quantum wires [60] as well as the electron-phonon phenomena in multilayer coaxial cylindrical $Al_xG1-xAs/GaAs$ quantum cables [61] and double coupled nanoshell systems [62]. The aim of the present work is to investigate feasibility of controlling the acoustic phonon energy spectra and corresponding phonon velocity dispersion in rolled-up micro- and nanoarchitectures. Of fundamental importance in this context is the experimental evidence [63], that due to oxide formation during fabrication, a single period of a radial superlattice is represented by a semiconductor/amorphous oxide/polycrystalline metal/amorphous oxide layer rather than a semiconductor/metal layer. It implies a necessity to investigate multilayer tubes. The elastodynamic boundary conditions on spiral interfaces of a rolled-up microtube with multiple windings or on cylindrical interfaces of a multilayer tube, which consists of coaxial cylindrical shells, (multishell) immediately affect acoustic phonon energy spectrum and, hence, phonon group velocities for propagation along the tube. Since the effect of these boundary conditions depends on the number of shells along with the geometric parameters, multishells are qualified into acoustic metamaterials. This introduces, in particular, extra capability for tuning the phonon spectrum, engineering the phonon transport and advancement of thermoelectric materials [9].

## 2. Theoretical Model

Spiral interfaces of a rolled-up nano- and microtube with multiple windings are modelled by cylindrical interfaces of a multilayer tube that consists of coaxial shells (multishell) as shown schematically in Fig. 1.

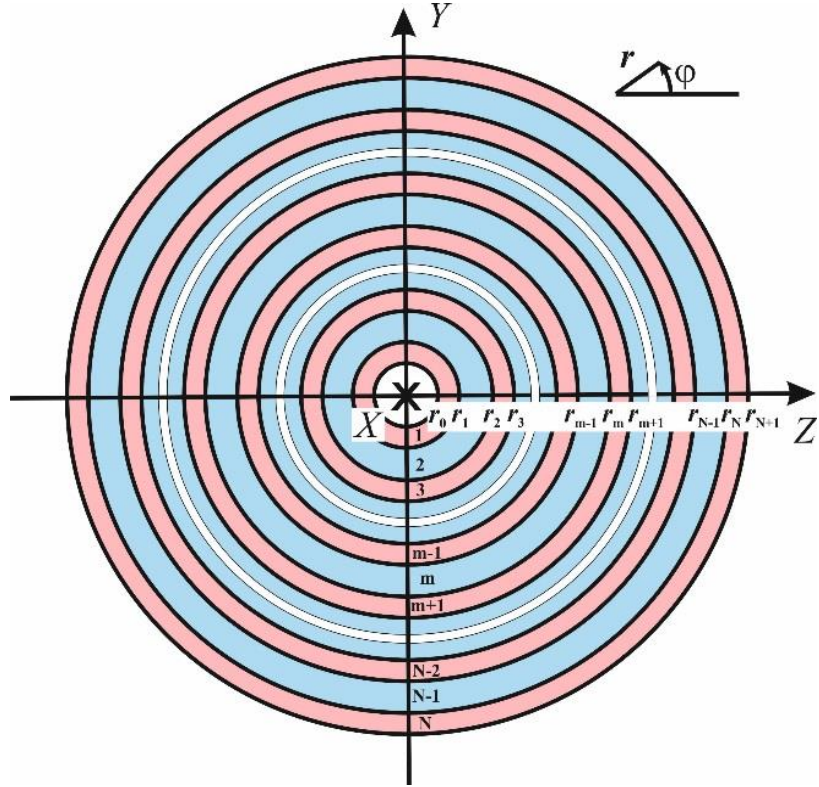

**Figure 1.** Cross-section in the plane orthogonal to the X-axis of a multilayer tube consisting of coaxial shells (multishell). The tube core and the outer medium are vacuum. The axis of the structure is selected as the X-axis. The polar coordinates in the YZ-plane are ($r,\varphi$). The picture corresponds to a multishell with a periodic alternation of two materials.

The displacement vector $\mathbf{u}_m$ in each layer ($m=1,..., N$) treated as an elastic continuum obeys the equations of elastodynamics [64]:

$$\mu_m \nabla^2 \mathbf{u}_m + (\lambda_m + \mu_m)\nabla\nabla\bullet\mathbf{u}_m = \rho_m \frac{\partial^2 \mathbf{u}_m}{\partial t^2}. \tag{1}$$

At every intershell boundary $r_m$, ($m=1,…, N$) the following six boundary conditions represent continuity of the stress tensor components

$$\begin{aligned}\sigma_{rrm} &= \sigma_{rrm+1};\\ \sigma_{rxm} &= \sigma_{rxm+1};\\ \sigma_{r\varphi m} &= \sigma_{r\varphi m+1};\end{aligned} \tag{2}$$

and of the displacement vector components

$$u_{rm} = u_{rm+1};$$
$$u_{xm} = u_{xm+1};$$
$$u_{\varphi m} = u_{\varphi m+1}.$$
(3)

At the internal boundary $r_0$, the following three boundary conditions represent vanishing of the stress tensor components

$$0 = \sigma_{rr1};$$
$$0 = \sigma_{r1};$$
$$0 = \sigma_{r\varphi 1}.$$
(4)

Similarly, at the external boundary $r_{N+1}$, the following three boundary conditions represent vanishing of the stress tensor components

$$\sigma_{rrN} = 0;$$
$$\sigma_{rxN} = 0;$$
$$\sigma_{r\varphi N} = 0.$$
(5)

In total, there are $6(N+1)$ boundary conditions. On that basis, the secular equation is derived for eigenmodes of phonons at an arbitrary number of the shell pairs. The boundary problem described by the Eq. (1) and the boundary conditions (2) to (5) satisfies the correspondence principle with respect to the case of a two-shell composite coaxial tube [65, 66].

The solutions to the equations of elastodynamics are sought in the form of a combination of dilatational waves and shear (equivoluminal) waves

$$\mathbf{u}_m = \Delta \Phi_m + \nabla \times \mathbf{H}_m.$$
(6)

Here the scalar potential of the dilatational motion and the vector potential of the shear motion along the axis of the structure (which has only two independent components) satisfy the equations:

$$v_{1m}^2 \nabla^2 \Phi_m = \ddot{\Phi}_m,$$
$$v_{2m}^2 \nabla^2 \mathbf{H}_m = \ddot{\mathbf{H}}_m.$$
(7)

The velocities $v_{1m}$ and $v_{2m}$ correspond to dilatational and shear waves in the material of the $m$-th layer. The solutions to the wave equations are sought as plane waves traveling along the axis of the structure.

Every neighboring layers are assumed to be perfectly bonded. Therefore, the eigenwaves in all layers will have the same longitudinal wave number $\xi$ and circular frequency $\omega$.

$$\Phi_m = f_m(r)\cos(n\varphi)\cos(\omega t + \xi x),$$
$$H_{rm} = h_{rm}(r)\sin(n\theta)\sin(\omega t + \xi x),$$
$$H_{\varphi m} = h_{\varphi m}(r)\cos(n\theta)\sin(\omega t + \xi x),$$
$$H_{xm} = h_{xm}(r)\sin(n\theta)\cos(\omega t + \xi x).$$
(8)

A multishell with a periodic alternation of two materials is further assumed (see Fig. 1, lower panel) with $r_0$=100 nm. All odd shells ($m=2k+1$) consist of one and the same material with elastic properties

$\lambda_1$, $\mu_1$, density $\rho_1$ and have the same thickness: $\Delta r_1$. All even shells ($m=2k$) consist of the same material with elastic properties $\lambda_2$, $\mu_2$, density $\rho_2$ and have the same thickness: $\Delta r_2$. They are represented in Table 1. In what follows, the number of layers is denoted by $N_L$.

**Table 1.** Geometric and materials parameters of the multishell

| Parity i of the layer number $m$ | 1 | 2 |
|---|---|---|
| Material | InAs | GaAs |
| $\lambda_i$, dyn/cm$^2$ | $4.54 \times 10^{11}$ | $5.34 \times 10^{11}$ |
| $\mu_i$, dyn/cm$^2$ | $1.90 \times 10^{11}$ | $3.285 \times 10^{11}$ |
| $\rho_i$, g/cm$^3$ | 5.68 | 5.317 |
| $\Delta r_i$, nm | 5 | 5 |

The wave characteristics in the materials are defined in Table 2.

**Table 2.** Definitions of the wave characteristics in the materials

| Characteristic of the layer with parity i of the number $m$ | |
|---|---|
| Velocity of a dilatational wave | $v_{1i} = [(\lambda_i + 2\mu_i)/\rho_i]^{1/2}$ |
| Velocity of a shear wave | $v_{2i} = [\mu_i/\rho_i]^{1/2}$ |
| Squared radial wave number for a dilatational wave | $\alpha_i^2 = \omega^2/v_{1i}^2 - \xi^2$ |
| Squared radial wave number for a shear wave | $\beta_i^2 = \omega^2/v_{2i}^2 - \xi^2$ |
| Dispersion of the Rayleigh waves $\alpha_i^2 = 0$ | $\omega = v_{1i}\xi$ |
| Dispersion of the Rayleigh waves $\beta_i^2 = 0$: | $\omega = v_{2i}\xi$ |

The phonon dispersion curves are represented in the non-dimensional form, using the units, which are defined in Table 3.

**Table 3.** Units for wave characteristics

| Physical quantity | Unit |
|---|---|
| Wave vector $\zeta$ | $1/\Delta r_2$ |
| Frequency $\omega$ | $\pi v_{22}/\Delta r_2$ |
| Group velocity $d\omega/d\zeta$ | $\pi v_{22}$ |

## 3. Results for $N_L=2$

The lowest phonon dispersion curves (in the window of eigenfrequencies [0, 1.5]) are shown for axially symmetric waves $n=0$, $N_L = 2$ in Fig. 2. (More time-consuming calculations for flexural waves with n=1,… are ongoing.) There are *anticrossings* of torsional or non-torsional modes, but there might occur *crossings* of torsional ($u_x=u_r=0$, $u_\theta \neq 0$) and non-torsional (associated with the displacement components $u_x$ and $u_r$, $u_\theta=0$) modes. Dispersion of the phonon group velocity for the dispersion curves in Fig. 2 is represented in Fig. 3 (see Fig. A1 for a detailed graph). A group velocity dispersion curves stops when the eigenfrequency $\omega$ goes beyond the window [0, 1.5].

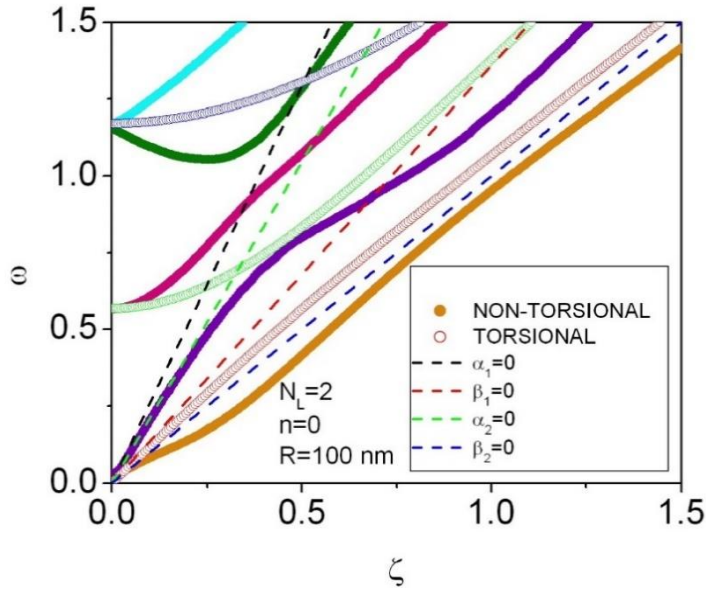

**Figure 2.** Phonon dispersion curves for $n=0$, $N_L=2$. Non-torsional and torsional modes are represented with filled and empty circles, correspondingly. Dashed lines indicate the dispersion curves for dilatational ($\alpha_i=0$) ans shear ($\beta_i=0$) waves in the material with parity $i$ of the number $m$.

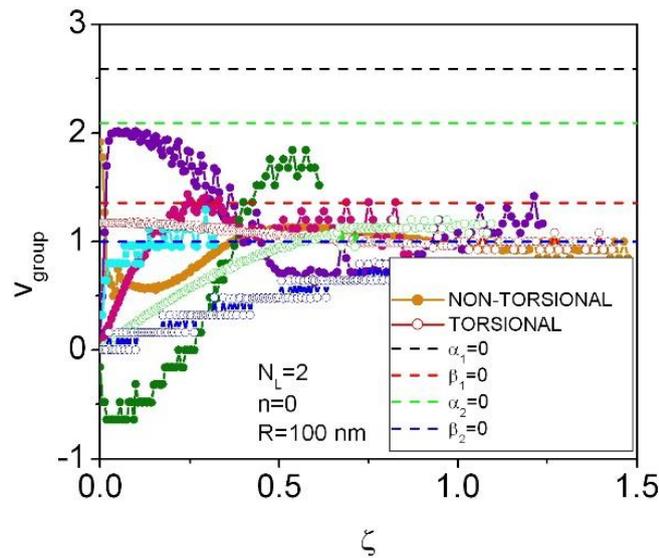

**Figure 3.** Phonon group velocity dispersion curves for $n=0$, $N_L=2$. Denotations of non-torsional and torsional modes are the same as in Fig. 2.

## 4. Results for $N_L$=4 and $N_L$=6

The lowest phonon dispersion curves are shown for axially symmetric waves $n$=0, $N_L$=4 in Fig. 4. For clarity, dispersion curves for non-torsional and torsional waves are represented also separately.

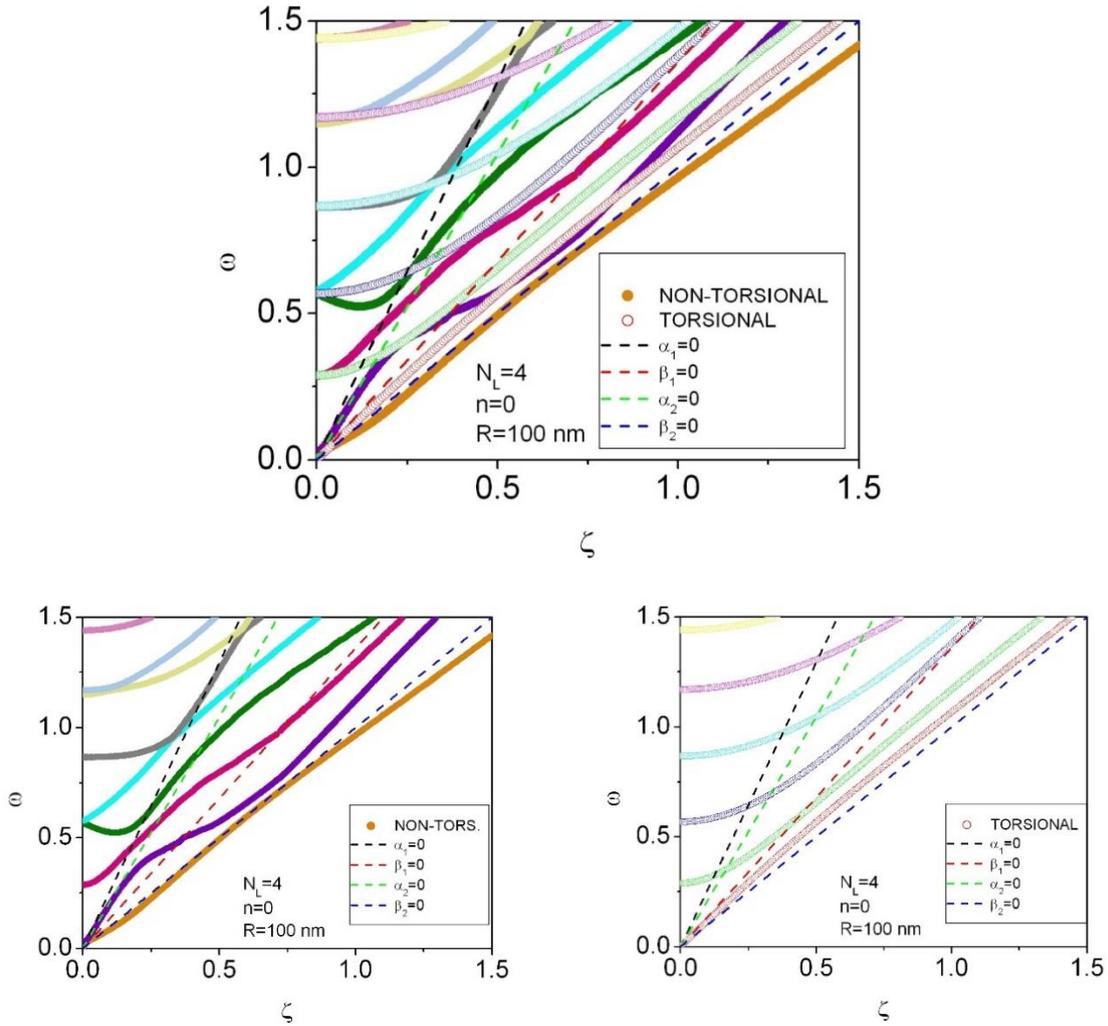

**Figure 4.** Phonon dispersion curves for $n$=0, $N_L$=4. Lower row: dispersion curves for non-torsional and torsional waves.

*A larger number of dispersion curves in multishells with 4 shells emerge within the same interval of energies and wave vectors as for multishells with 2 shells in Fig. 2.* Dispersion of the phonon group velocity for the lowest dispersion curves in Fig. 4 is represented in Fig. 5 (see Fig. A2 for a detailed graph).

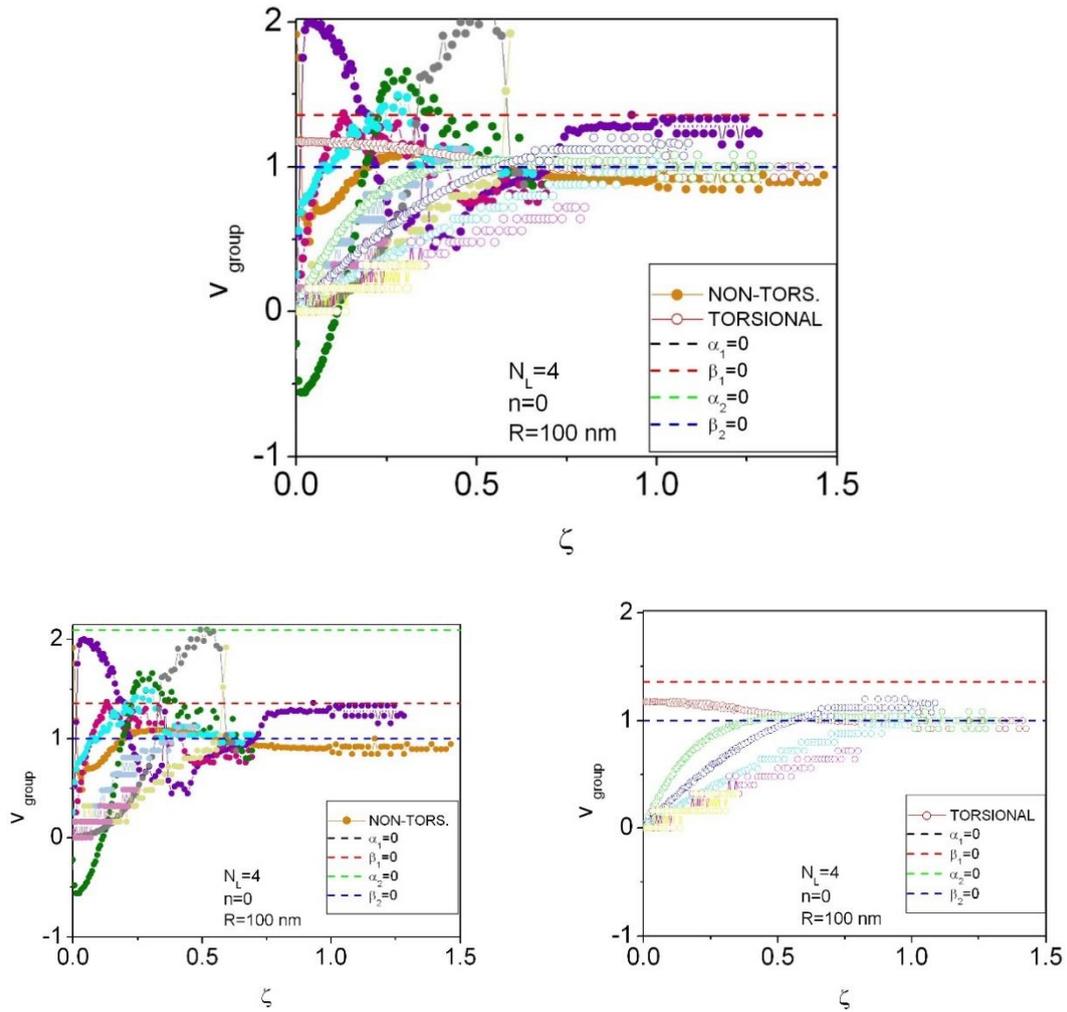

**Figure 5.** Phonon group velocity dispersion curves for $n=0$, $N_L=4$.

The lowest phonon dispersion curves are shown for axially symmetric waves $n=0$, $N_L=6$ in Fig. 6.

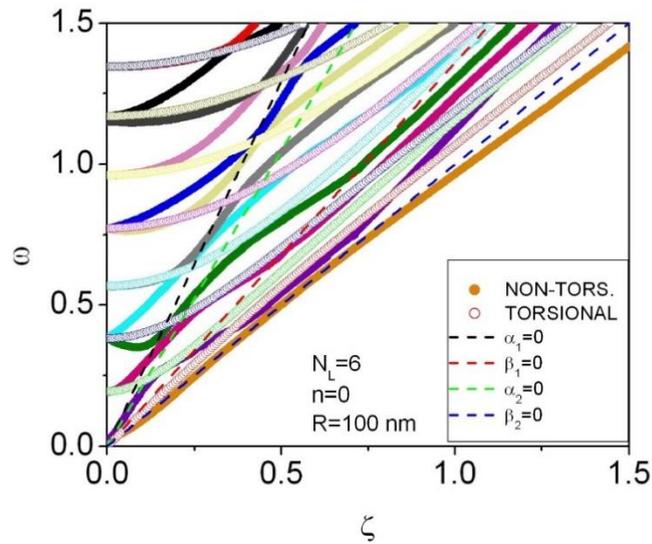

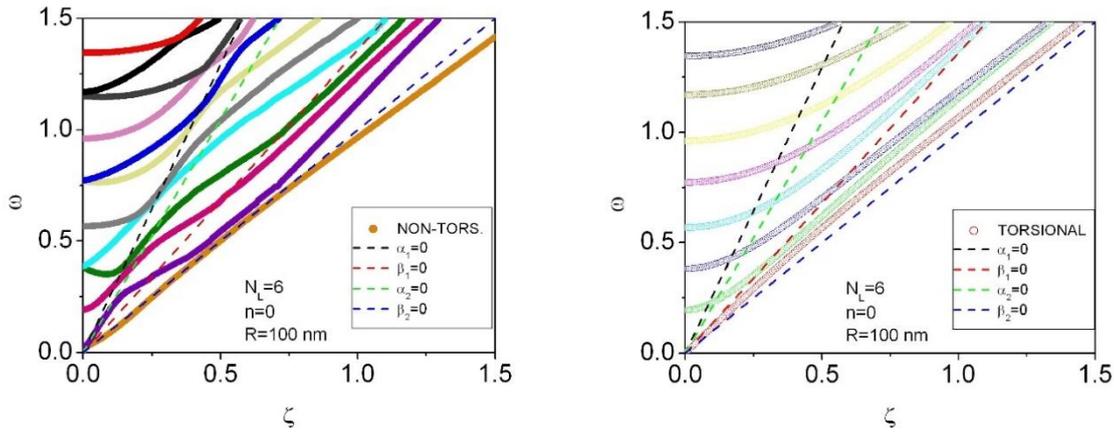

**Figure 6.** Phonon dispersion curves for $n=0$, $N_L=6$.

A larger number of dispersion curves in multishells with 6 shells emerge within the same interval of energies and wave vectors as for multishells with 4 shells in Fig. 4 and even more so for multishells with 2 shells in Fig. 2. Dispersion of the phonon group velocity for the lowest dispersion curves in Fig. 6 is represented in Fig. 7 (see Fig. A3 for a detailed graph).

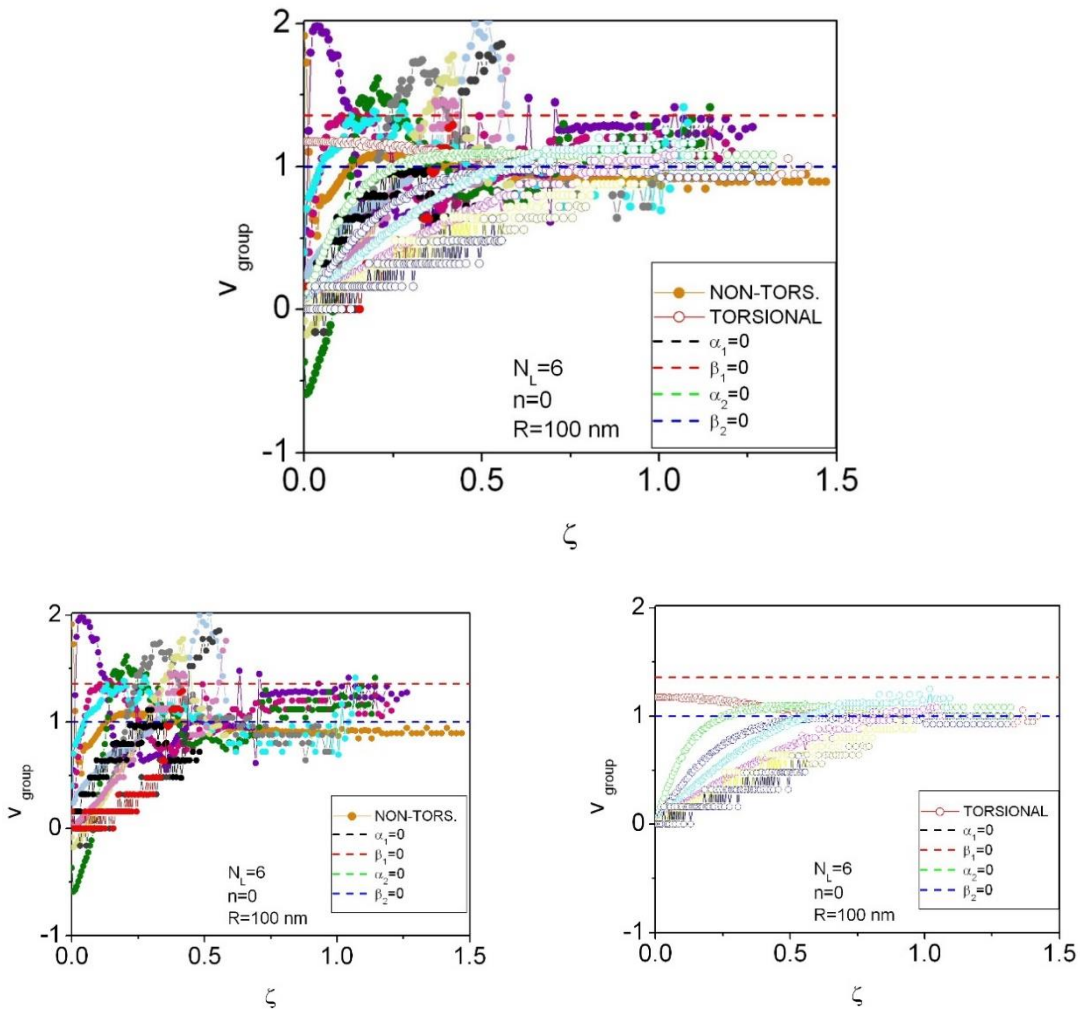

**Figure 7.** Phonon group velocity dispersion curves for $n=0$, $N_L=6$.

## 5. Geometric Effects in the Phonon Dispersion and Group Velocities for Different Numbers of Layers

For the axially symmetric waves ($n=0$), as follows from Figs. 2, 4 and 6, the lowest group of the phonon dispersion curves, containing one torsional and two non-torsional modes, at the small wave vectors $\zeta$ is only slighltly changed by the number of layers $N_L$. In the same region of wave vectors, the second, consisting of one torsional and one non-torsional modes (the third, consisting of one torsional and two non-torsional modes) group of the phonon frequencies ω significantly decreases from 0.57 (1.17) for $N_L=2$ to 0.29 (0.57) for $N_L=4$ and 0.19 (0.38) for $N_L=6$. Within the numerical accuracy, the decrease of the phonon frequencies in the long-wave limit is *inversely proportional* to $N_L$. Away from the long-wave limit, a general trend of "compression" of the phonon energy spectrum towards lower values of phonon frequencies persists.

As seen from Figs. 3, 5 and 7, the phonon group velocity related to the fundamental (lowest) torsional mode is a weakly varying function of the wave vector $\zeta$, while for the higher torsional modes it monotonously increases with the wave vector $\zeta$, apparently towards a saturation. For a fixed value $\zeta=0.05$, the phonon group velocity related to the lowest (second lowest) torsional mode depends on the number of layers $N_L$ as follows: 1.17 (0.13) for $N_L=2$; 1.17 (0.27) for $N_L=4$; 1.17 (0.37) for $N_L=6$. Within the numerical accuracy, the increase of the phonon group velocity for the second lowest torsional mode is *directly propotional* to $N_L$.

The phonon group velocity related to the lowest two non-torsional modes is a weak function of the wave vector $\zeta$, while for higher torsional modes it always strongly depends on $\zeta$. For the same fixed value $\zeta=0.05$ as above, the phonon group velocity related to the lowest (second lowest and third lowest) non-torsional mode depends on the number of layers $N_L$ as follows: 0.75 (2.02 and 0.43) for $N_L=2$; 0.75 (1.98 and 0.82) for $N_L=4$; 0.78 (1.94 and 1.06) for $N_L=6$. Within the numerical accuracy, the phonon group velocity for the third lowest torsional mode reveals a *sublinear dependence* on $N_L$.

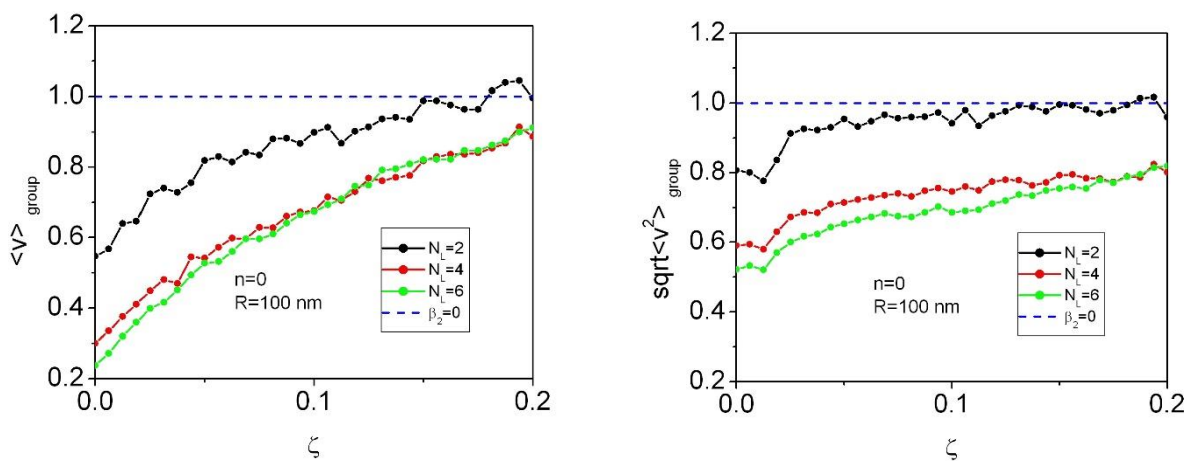

**Figure 8.** Average (**left**) and rms (**right panel**) phonon group velocity dispersion curves for $n=0$ at $N_L = 2, 4, 6$.

Finally, in order to clearly vizualize the overall impact of the number of layers in a multishell on the group velocity dispersion, the average and rms phonon group velocities are calculated for the branches available from the results of the previous section. The results, shown in Fig. 8, demonstrate that ***an*** *increase of $N_L$ from 2 to 4 leads to an appreciable decrease of the average and rms phonon group velocities*. A further increase of $N_L$ from 4 to 6 has a smaller impact on the average and rms phonon group velocities. For the wave vector $\zeta=0.05$, the average phonon group velocity decreases from 0.82 for $N_L=2$ to 0.54 for $N_L=4$ and further to 0.53 for $N_L=6$. At the same time, the rms phonon group velocity is reduced from 0.95 for $N_L=2$ to 0.71 for $N_L=4$ and further to to 0.65 for $N_L=6$. At small wave vectors the trend persists: the average and rms phonon group velocities decrease with increasing $N_L$.

## 6. Conclusions

We established a possibility of efficient engineering of the acoustic phonon energy dispersion in multishell tubular structures produced by a novel method of self-assembly of micro- and nano-architectures. A dependence on the number of layers in a multishell structure is a manifestation of geometric effects on phonon energy spectrum. Such geometric effects are features pertinent to acoustic metamaterials and phonon crystals. Based on the calculated energies, the phonon confinement effects should be directly observable using Brillouin spectrometry. The changes in the acoustic phonon spectrum affect phonon transport and can be experimentally detected in thermal conductivity measurements. The reduction of the phonon group velocity and phonon thermal conductivity can be achieved without significant roughness scattering and degradation of electron transport. Our results suggest that arrays of rolled-up multishell tubular structures are prospective candidates for advancement in thermoelectric materials and devices.

**Appendix. Dispersion of the Phonon Group Velocity**

Figures A1 to A3 represent the detailed phonon group velocity for the wavevector window [0, 0.5]. Step-like features provide a measure of precision for the exploited numerical procedure of solving the boundary problem of Eqs. (1) to (5).

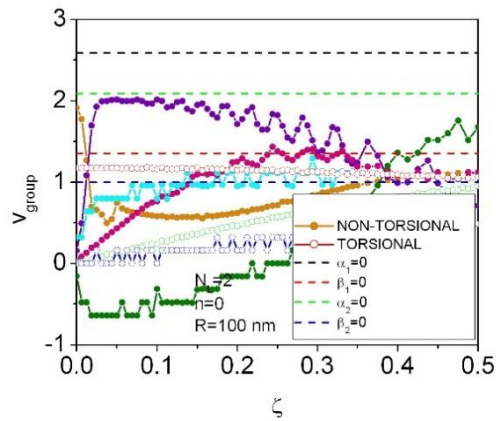

**Figure A1.** Detailed phonon group velocity dispersion curves for $n=0$, $N_L=2$. Denotations of non-torsional and torsional modes are the same as in Fig. 2.

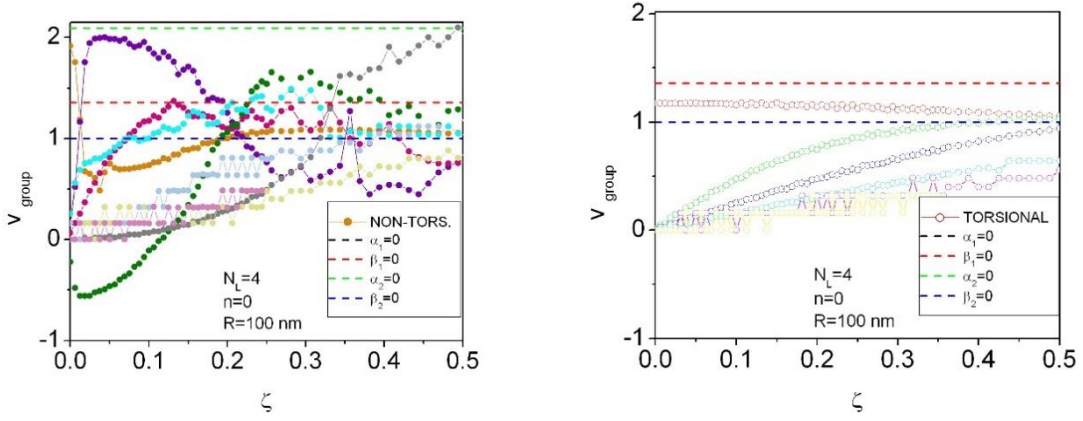

**Figure A2.** Detailed phonon group velocity dispersion curves for $n$=0, $N_L$=4.

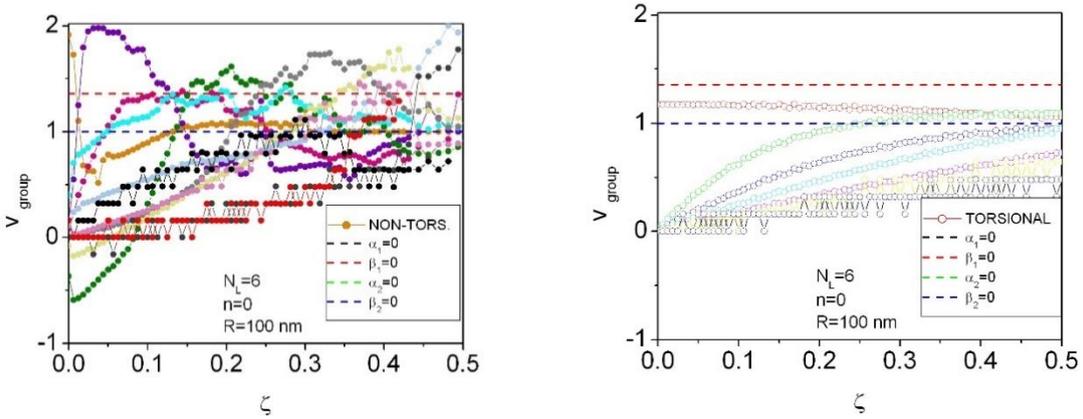

**Figure A3.** Detailed phonon group velocity dispersion curves for $n$=0, $N_L$=6.

## Acknowledgments

Discussions with O. G. Schmidt are gratefully acknowledged. The work at the IIN IFW Dresden was partly supported by the Deutsche Forschungsgemeinschaft (DFG) under Grant # FO 956/2-1. The work at UC Riverside was supported as part of the Spins and Heat in Nanoscale Electronic Systems (SHINES), an Energy Frontier Research Center funded by the U.S. Department of Energy, Office of Science, Basic Energy Sciences (BES) under Award # SC0012670.

## Author Contributions

The authors contributed equally to this work.

## Conflicts of Interest

The authors declare no conflict of interest.

**References and Notes**

1. Fomin, V.M.; Pokatilov, E.P. Phonons and the electron-phonon interaction in multi-layer systems. Phys. Stat. Sol. (b) 1985, 132, 69-82.
2. Pokatilov, E.P.; Fomin, V.M.; Beril, S.I. *Vibrational excitations, polarons and excitons in multi-layer structures and superlattices*; Shtiintsa: Kishinev, 1990; 280 pp.
3. Pokatilov, E.P.; Fomin, V.M.; Semenovskaya, N.N.; Beril, S.I. Interaction Hamiltonian between an electron and polar surface vibrations in a symmetrical three-layer structure. Phys. Rev. B 1993, 47, 16597-16600.
4. Klimin, S.N.; Pokatilov, E.P.; Fomin, V.M. Bulk and interface polarons in quantum wires and dots. Phys. Stat. Sol.(b) 1994, 184, 373-383.
5. Pokatilov, E.P.; Fomin, V.M.; Klimin, S.N.; Balaban, S.N. Characterization of nanostructures by virtue of the phenomena due to the electron-phonon interaction. Applied Surface Science 1996, 104/105, 546-551.
6. Stroscio, M.A.; Dutta, M. *Phonons in Nanostructures*; Cambridge University Press: Cambridge, 2001; 282 pp.
7. Balandin, A.A. Nanophononics: Phonon engineering in nanostructures and nanodevices. J. Nanosci. Nanotech. 2005, 5, 1015-1022.
8. Balandin, A.A.; Pokatilov, E.P.; Nika, D.L. Phonon engineering in hetero- and nanostructures. J. Nanoelectron. Optoelectron. 2007, 2, 140-170.
9. Balandin, A.A.; Nika, D.L. Phononics in low-dimensional materials. Materials Today 2012, 15, 266–275.
10. Balandin A.; Wang, K.L. Significant decrease of the lattice thermal conductivity due to phonon confinement in a free-standing semiconductor quantum well. Phys. Rev. B 1998, 58, 1544-1549.
11. Pokatilov, E.P.; Nika, D.L.; Balandin, A.A. Phonon spectrum and group velocities in AlN/GaN/AlN and related heterostructures. Superlatt. Microstruct. 2003, 33, 155-171.
12. Mingo, N.; Yang, L.; Li, D.; Majumdar, A. Predicting the thermal conductivity of Si and Ge nanowires. Nano Lett. 2003, 3, 1713-1716.
13. Mingo, N. Calculation of Si nanowire thermal conductivity using complete phonon dispersion relations. Phys. Rev. B 2003, 68, 113308, 1-4.
14. Pokatilov, E.P.; Nika, D.L.; Balandin, A.A. Acoustic-phonon propagation in rectangular semiconductor nanowires with elastically dissimilar barriers. Phys. Rev. B 2005, 72, 113311, 1-4.
15. Bannov, N.; Aristov, V.; Mitin, V.; Stroscio, M.A. Electron relaxation times due to the deformation-potential interaction of electrons with confined acoustic phonons in a free-standing quantum well. Phys. Rev. B 1995, 51, 9930-9942.
16. Stroscio, M. A. Interaction between longitudinal-optical-phonon modes of a rectangular quantum wire and charge carriers of a one-dimensional electron gas. Phys. Rev. B 1989, 40, 6428-6431.
17. Mickevičius, R.; Mitin, V. Acoustic-phonon scattering in a rectangular quantum wire. Phys. Rev. B 1993, 48, 17194-17201.
18. Mickevičius, R.; Mitin, V.; Harithsa, U.K.; Jovanovic D.; Leburton, J.P. Superlinear electron transport and noise in quantum wires. J. Appl. Phys. 1994, 75, 973-978.

19. Svizhenko, A.; Balandin, A.; Bandyopadhyay, S.; Stroscio, M.A. Electron interaction with confined acoustic phonons in quantum wires subjected to a magnetic field. Phys. Rev. B 1998, 57, 4687-4693.
20. Svizhenko, A.; Bandyopadhyay, S.; Stroscio, M.A. The effect of acoustic phonon confinement on the momentum and energy relaxation of hot carriers in quantum wires. J. Phys.: Cond. Matter 1998, 10, 6091-6094.
21. Pokatilov, E.P.; Nika, D.L.; Balandin, A.A. Confined electron-confined phonon scattering rates in wurtzite AlN/GaN/AlN heterostructures. J. Appl. Phys. 2004, 95, 5626-5632.
22. Fomin, V.M.; Pokatilov, E.P.; Devreese, J.T.; Klimin, S.N.; Gladilin, V.N.; Balaban, S.N. Multiquantum optical processes in semiconductor quantum dots. Phys. Stat. Sol. (b) 1997, 164, 417-420.
23. Fomin, V.M.; Gladilin, V.N.; Devreese, J.T.; Pokatilov, E.P.; Balaban, S.N.; Klimin, S.N. Photoluminescence of spherical quantum dots. Phys. Rev. B 1998, 57, 2415-2425.
24. Fomin, V.M.; Klimin, S.N.; Gladilin, V.N.; Devreese, J.T. Characterization of self-assembled quantum dots using the phonon-induced features of PL spectra. J. Lumin. 2000, 87-89, 330-332.
25. Fomin, V.M.; Devreese, J.T. Theory of excitons in semiconductor quantum dots. Nonlinear Optics 2002, 29, 321-327.
26. Fonoberov, V.A.; Pokatilov, E.P.; Fomin, V.M.; Devreese, J.T. Photoluminescence of tetrahedral quantum-dot quantum wells. Phys. Rev. Lett. 2004, 92, 127402, 1-4.
27. Klimin, S.N.; Fomin, V.M.; Devreese, J.T.; Bimberg, D. Model of Raman scattering in self-assembled InAs/GaAs quantum dots. Phys. Rev. B 2008, 77, 045307, 1-11.
28. Khurgin, J.B.; Sun, G. Enhancement of light absorption in a quantum well by surface plasmon polariton. Appl. Phys. Lett. 2009, 94, 191106, 1-3.
29. Balandin, A.; Wang, K.L. Effect of phonon confinement on the thermoelectric figure of merit of quantum wells. J. Appl. Phys. 1998, 84, 6149-6153.
30. Khitun, A.; Balandin, A.; Wang, K.L. Modification of the lattice thermal conductivity in silicon quantum wires due to spatial confinement of acoustic phonons. Superlatt. Microstruct. 1999, 26, 181-193.
31. Zou, J.; Balandin, A. Phonon heat conduction in a semiconductor nanowire. J. Appl. Phys. 2001, 89, 2932-2938.
32. Fonoberov, V.A.; Balandin, A.A. Giant enhancement of the carrier mobility in Silicon nanowires with diamond coating. Nano Lett. 2006, 6, 2442-2446.
33. Nika, D.L.; Pokatilov, E.P.; Balandin, A.A. Phonon-engineered mobility enhancement in the acoustically mismatched silicon/diamond transistor channels. Appl. Phys. Lett. 2009, 93, 173111, 1-3.
34. Zincenco, N.D.; Nika, D.L.; Pokatilov, E.P.; Balandin, A.A. Acoustic phonon engineering of thermal properties of silicon-based nanostructures. J. Phys. Conf. 2007, Series 92, 012086, 1-4.
35. Wingert, M.C.; Chen, Z.C.Y.; Dechaumphai, E.; Moon, J.; Kim, J.-H.; Xiang, J.; Chen, R. Thermal conductivity of Ge and Ge–Si core–shell nanowires in the phonon confinement regime. Nano Lett. 2011, 11, 5507–5513.
36. Cuffe, J.; Chávez, E.; Shchepetov, A.; Chapuis, P.-O.; El Boudouti E.H.; Alzina, F.; Kehoe, T.; Gomis-Bresco, J.; Dudek, D. ; Pennec, Y.; Djafari-Rouhani, B.; Prunnila, M.; Ahopelto, J.; Torres,

C.M.S. Phonons in slow motion: Dispersion relations in ultrathin Si membranes. Nano Lett. 2012, 12, 3569–3573.
37. Johnson, W.L.; Kim, S.A.; Geiss, R.; Flannery, C.M.; Bertness, K.A.; Heyliger, P.R. Vibrational modes of GaN nanowires in the gigahertz range. Nanotechnology 2012, 23, 495709, 1-11.
38. Pokatilov, E.P.; Nika, D.L.; Balandin, A.A. A phonon depletion effect in ultrathin heterostructures with acoustically mismatched layers. Appl. Phys. Lett. 2004, 85, 825-827.
39. Nika, D.L.; Zincenco, N.D.; Pokatilov, E.P. Engineering of thermal fluxes in phonon mismatched heterostructures. J. Nanoelect. Optoelect. 2009, 4, 180-185.
40. Hu, M.; Giapis, K.P.; Goicochea, J.V.; Zhang, X.; Poulikakos, D. Significant reduction of thermal conductivity in Si/Ge core−shell nanowires. Nano Lett. 2011, 11, 618–623.
41. Bi, K.; Wang, J.; Wang, Y.; Sha, J.; Wang, Z.; Chen, M.; Chen, Y. The thermal conductivity of SiGe heterostructure nanowires with different cores and shells. Phys. Lett. A 2012, 376, 2668–2671.
42. He, Y.; Galli, G. Microscopic origin of the reduced thermal conductivity of Silicon nanowires. Phys. Rev. Lett. 2012, 108, 215901, 1-5.
43. Petrin, A., Ed. *Wave Propagation in Materials for Modern Applications*; InTech: Rijeka, Croatia, 2010; ISBN 978-953-7619-65-7; 552 pp.
44. Deymier, P.A., Ed. *Acoustic Metamaterials and Phononic Crystals*; Springer: Berlin☐Heidelberg, 2013; 378 pp.
45. Narayana, S.; Sato, Y. Heat flux manipulation by engineered thermal materials. Phys. Rev. Lett. 2012, 108, 214303, 1-5.
46. Yu, B.; Yang, S.; Gan, C.; Lei, H. A new procedure for exploring the dispersion characteristics of longitudinal guided waves in a multi-layered tube with a weak interface. J. Nondestruct. Eval. 2013, 32, 263–276.
47. Wang, L.; Hu, H.; Guo, W. Wave Propagation in Carbon Nanotubes. In Ref. [43], 225-252.
48. Sigalas, M.M.; Economou, M.N. Elastic and acoustic wave band structure. Journal of Sound and Vibrations 1992, 158, 377-382.
49. Sigalas, M.M.; Economou, M.N. Band structure of elastic waves in two dimensional systems. Solid State Communs. 1993, 86, 141-143.
50. Kushwaha, M.S.; Halevi, P.; Dobrzynski, L.; Djafari-Rouhani, B. Acoustic band structure of periodic elastic composites. Phys. Rev. Lett. 1993, 71, 2022–2025.
51. Liu, Z.; Zhang, X.; Mao, Y.; Zhu, Y.Y.; Yang, Z.; Chan, C.T.; Sheng, P. Locally resonant sonic materials. Science 2000, 289, 1734-1736.
52. Movchan, A.B.; Guenneau, S. Split-ring resonators and localized modes. Phys. Rev. B 2004, 70, 125116, 1-5.
53. Li, J.; Chan, C.T. Double-negative acoustic metamaterial. Phys. Rev. E 2004, 70, 055602(R), 1-4.
54. Prinz, V.Ya.; Seleznev, V.A.; Gutakovsky, A.K.; Chehovskiy, A.V.; Preobrazhenskii, V.V.; Putyato, M.A.; Gavrilova, T.A. Free-standing and overgrown InGaAs/GaAs nanotubes, nanohelices and their arrays. Physica E 2000, 6, 828-831.
55. Schmidt, O.G.; Eberl, K. Nanotechnology—thin solid films roll up into nanotubes. Nature 2001, 410, 168.

56. Krause, B.; Mocuta, C.; Metzger, T.H.; Deneke, C.; Schmidt, O.G. Local structure of a rolled-up single crystal: An X-ray microdiffraction study of individual semiconductor nanotubes. Phys. Rev. Lett. 2006, 96, 165502, 1-4.
57. Deneke, C.; Jin-Phillipp, N.-Y.; Loa, I.; Schmidt, O.G. Radial superlattices and single nanoreactors. Appl. Phys. Lett. 2004, 84, 4475−4477.
58. Deneke, C.; Songmuang, R.; Jin-Phillipp, N.Y.; Schmidt, O.G. The structure of hybrid radial superlattices. J. Phys. D: Appl. Phys. 2009, 42, 103001, 1-16.
59. Grimm, D.; Wilson, R.B.; Teshome, B.; Gorantla, S.; Rümmeli, M.H.; Bublat, T.; Zallo, E.; Li, G.; Cahill, D.G.; Schmidt, O.G. Thermal conductivity of mechanically joined semiconducting/metal nanomembrane superlattices. Nano Lett. 2014, 14, 2387−2393.
60. Pokatilov, E.P.; Fomin, V.M.; Devreese, J.T.; Balaban, S.N.; Klimin, S.N. Bipolaron binding in quantum wires. Phys. Rev. B 2000, 61, 2721-2728.
61. Zhang, L.; Xie, H.-J. Fröhlich electron-interface and -surface optical phonon interaction Hamiltonian in multilayer coaxial cylindrical $Al_xGa_{1-x}As/GaAs$ quantum cables. Journal of Physics: Condensed Matter 2003, 15, 5871-5879.
62. Kanyinda-Malu, C.; Clares F.J.; de la Cruz, R.M. Axial interface optical phonon modes in a double-nanoshell system. Nanotechnology 2008, 19, 285713, 1-8.
63. Deneke, C.; Sigle, W.; Eigenthaler, U.; van Aken, P.A.; Schütz G.; Schmidt, O.G. Interfaces in semiconductor/metal radial superlattices. Appl. Phys. Lett. 2007, 90, 263107, 1-3.
64. Graff, K.F. *Wave Motion in Elastic Solids*; Dover: New York, 1991; 688 pp.
65. Armenàkas, A.E. Propagation of Harmonic Waves in Composite Circular Cylindrical Shells. I: Theoretical investigation. American Institute of Aeronautics and Astronautics (AIAA) Journal 1966, 5, 740-744.
66. Armenàkas, A.E. Propagation of Harmonic Waves in Composite Circular Cylindrical Shells. Part II: Numerical Analysis. American Institute of Aeronautics and Astronautics (AIAA) Journal 1971, 9, 599-605.